\documentclass[11pt]{article}
\usepackage{amsmath,epsfig,sint}
\usepackage[american]{babel}
\font\sixrm=cmr6

\def\rmO{{\rm O}}

\def\bfy{{\bf y}}
\def\bfz{{\bf z}}

\def\proof{\noindent{\sl Proof:}\kern0.6em}

\def\frac#1#2{\hbox{$#1\over#2$}}
\def\dual{\mathstrut^*\kern-0.1em}

\def\lvec#1{\setbox0=\hbox{$#1$}
    \setbox1=\hbox{$\scriptstyle\leftarrow$}
    #1\kern-\wd0\smash{
    \raise\ht0\hbox{$\raise1pt\hbox{$\scriptstyle\leftarrow$}$}}
    \kern-\wd1\kern\wd0}
\def\rvec#1{\setbox0=\hbox{$#1$}
    \setbox1=\hbox{$\scriptstyle\rightarrow$}
    #1\kern-\wd0\smash{
    \raise\ht0\hbox{$\raise1pt\hbox{$\scriptstyle\rightarrow$}$}}
    \kern-\wd1\kern\wd0}

\def\nabstar#1{\nabla\kern-0.5pt\smash{\raise 4.5pt\hbox{$\ast$}}
               \kern-4.5pt_{#1}}

\def\drvstar#1{\partial\kern-0.5pt\smash{\raise 4.5pt\hbox{$\ast$}}
               \kern-5.0pt_{#1}}

\def\momp#1#2{
    \setbox0=\hbox{${#1}$}\setbox1=\hbox{${#1}_{#2}$}
    {#1}_{#2}\kern-\wd1\kern\wd0
    \smash{\raise4.5pt\hbox{$\scriptscriptstyle +$}}}
\def\momm#1#2{
    \setbox0=\hbox{${#1}$}\setbox1=\hbox{${#1}_{#2}$}
    {#1}_{#2}\kern-\wd1\kern\wd0
    \smash{\raise4.5pt\hbox{$\scriptscriptstyle -$}}}
\def\mompm#1#2{
    \setbox0=\hbox{${#1}$}\setbox1=\hbox{${#1}_{#2}$}
    {#1}_{#2}\kern-\wd1\kern\wd0
    \smash{\raise4.5pt\hbox{$\scriptscriptstyle \pm$}}}
\def\smomp#1#2{
    \setbox0=\hbox{${#1}$}\setbox1=\hbox{${#1}_{#2}$}
    {#1}_{#2}\kern-\wd1\kern\wd0
    \smash{\raise3pt\hbox{$\scriptscriptstyle +$}}}
\def\smomm#1#2{
    \setbox0=\hbox{${#1}$}\setbox1=\hbox{${#1}_{#2}$}
    {#1}_{#2}\kern-\wd1\kern\wd0
    \smash{\raise3pt\hbox{$\scriptscriptstyle -$}}}
\def\smompm#1#2{
    \setbox0=\hbox{${#1}$}\setbox1=\hbox{${#1}_{#2}$}
    {#1}_{#2}\kern-\wd1\kern\wd0
    \smash{\raise3pt\hbox{$\scriptscriptstyle \pm$}}}
\def\si{\kern1pt{\rm si}}
\def\co{\kern1pt{\rm co}}

\def\Nf{N_{\rm f}}
\def\psibar{\bar{\psi}}

\def\psiprime{\psi\kern1pt'}
\def\psibarprime{\psibar\kern1pt'}
\def\rhoprime{\rho\kern1pt'}
\def\rhobar{\bar{\rho}}
\def\rhobarprime{\rhobar\kern1pt'}
\def\rhobartilde{\kern2pt\tilde{\kern-2pt\rhobar}}
\def\rhobartildeprime{\kern2pt\tilde{\kern-2pt\rhobar}\kern1pt'}

\def\zetabar{\bar{\zeta}}
\def\zetaprime{\zeta\kern1pt'}
\def\zetabarprime{\zetabar\kern1pt'}
\def\zetar{\zeta_{\raise-1pt\hbox{\sixrm R}}}
\def\zetabarr{\zetabar_{\raise-1pt\hbox{\sixrm R}}}

\def\phiimpr{\phi_{\kern0.5pt\hbox{\sixrm I}}}

\def\diracstar#1#2{
    \setbox0=\hbox{$\gamma$}\setbox1=\hbox{$\gamma_{#1}$}
    \gamma_{#1}\kern-\wd1\kern\wd0
    \smash{\raise4.5pt\hbox{$\scriptstyle#2$}}}

\def\Rap{R_{\rm AP}}
\def\Rm{R_{\rm m}}

\def\ba{b_{\rm A}}

\def\bp{b_{\rm P}}

\def\bv{b_{\rm V}}

\def\bg{b_{\rm g}}
\def\bm{b_{\rm m}}

\def\ca{c_{\rm A}}
\def\cv{c_{\rm V}}

\def\csw{c_{\rm sw}}

\def\fa{f_{\rm A}}

\def\fp{f_{\rm P}}

\def\f1{f_1}

\def\h1{h_1}

\def\opprime#1{\setbox0=\hbox{${\cal O}$}\setbox1=\hbox{${\cal O}_{\rm #1}$}
    {\cal O}_{\rm #1}\kern-\wd1\kern\wd0
    \smash{\raise4.5pt\hbox{\kern1pt$\scriptstyle\prime$}}\kern1pt}

\def\ophatprime#1{\setbox0=\hbox{$\widehat{\cal O}$}
    \setbox1=\hbox{$\widehat{\cal O}_{\rm #1}$}
    \widehat{\cal O}_{\rm #1}\kern-\wd1\kern\wd0
    \smash{\raise4.5pt\hbox{\kern1pt$\scriptstyle\prime$}}\kern1pt}

\def\bopprime#1{\setbox0=\hbox{${\cal O}$}\setbox1=\hbox{${\cal O}_{\rm #1}$}
    {\cal L}_{\rm #1}\kern-\wd1\kern\wd0
    \smash{\raise4.5pt\hbox{\kern1pt$\scriptstyle\prime$}}\kern1pt}

\def\blagprime#1{\setbox0=\hbox{${\cal B}$}\setbox1=\hbox{${\cal B}_{#1}$}
    {\cal B}_{#1}\kern-\wd1\kern\wd0
    \smash{\raise5.2pt\hbox{\kern1pt$\scriptstyle\prime$}}\kern1pt}

\def\mq{m_{\rm q}}
\def\mqtilde{\widetilde{m}_{\rm q}}

\def\mr{m_{{\hbox{\sixrm R}}}}

\def\mc{m_{\rm c}}

\def\za{Z_{\rm A}}

\def\zp{Z_{\rm P}}

\def\zm{Z_{\rm m}}

\def\Za{\za}

\def\Zp{\zp}

\def\Zm{\zm}

\def\gtilde{\tilde{g}_0}

\def\msbar{{\rm \overline{MS\kern-0.05em}\kern0.05em}}

\newcommand{\bes}{\begin{eqnarray}}
\newcommand{\ees}{\end{eqnarray}}

\begin{document}
\begin{titlepage}
\begin{flushright}
   DESY 00-131\\
   HUB-EP-2000/32\\
   ROM2F/2000/28\\
   September 2000
\end{flushright}
\vskip 0.5 cm
\begin{center}
 {\Large\bf  Non-perturbative results for the coefficients \\[1ex]
   $\bm$ and $\ba-\bp$ in O($a$) improved lattice QCD\\[1.5ex]
    }
\end{center}
\vskip 0.5 cm
\begin{figure}[h]
\begin{center}
\epsfig{figure=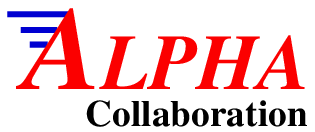} 
\end{center}
\end{figure}
\begin{center}
{\large Marco Guagnelli$^{\scriptscriptstyle a}$,
        Roberto Petronzio$^{\scriptscriptstyle a}$,
        Juri Rolf$^{\scriptscriptstyle \,b}$,\\[1ex]
        Stefan Sint$^{\scriptscriptstyle a}$,
        Rainer Sommer$^{\scriptscriptstyle c}$
    and Ulli Wolff$^{\scriptscriptstyle \,b}$}
\end{center}
\vskip 2.3ex
\begin{flushleft}
$^{\scriptstyle a}$ Universit\`a di Roma ``Tor Vergata'',
Dipartimento di Fisica, \\  Via della Ricerca Scientifica 1, 
I-00133 Rome, Italy\\[1ex]
$^{\scriptstyle b}$ Institut f\"ur Physik, Humboldt Universit\"at, 
Invalidenstr. 110, \\ D-10099 Berlin, Germany\\[1ex]
$^{\scriptstyle c}$ DESY, Platanenallee 6, D-15738 Zeuthen, Germany
\end{flushleft}
\begin{center} 
 {\bf Abstract}
\end{center}
\vskip 0.7ex
We determine the improvement coefficients $\bm$ and
$\ba-\bp$ in quenched lattice QCD for a range of $\beta$-values, which is
relevant for current large scale simulations. At fixed $\beta$, 
the results are rather sensitive to the precise choices of parameters. 
We therefore impose improvement conditions at constant 
renormalized parameters, and the coefficients are then obtained as smooth functions
of $g_0^2$. Other improvement conditions yield a different functional
dependence, but the difference between the coefficients
vanishes with a rate  proportional to the lattice spacing.
We verify this theoretical expectation in a few examples and
are therefore confident that O($a$) improvement is achieved 
for physical quantities. As a byproduct of our analysis 
we also obtain the finite renormalization constant which 
relates the subtracted bare quark mass to the bare PCAC mass.
\vfill
\eject

\vfill

\eject

\end{titlepage}

\section{Introduction}

Lattice QCD with Wilson quarks provides an attractive framework for
studying the strong interactions beyond perturbation theory. 
Its main shortcoming consists in the explicit breaking of all axial
symmetries. This is generally not considered a fundamental problem,
but it renders the renormalization of the theory more complicated.
In addition, the lack of chiral symmetry is at the origin of many O($a$) 
lattice artifacts in physical quantities ($a$ is the lattice spacing),
which can be rather large in practice~\cite{letter,aeffects}. 
In order to ease continuum extrapolations it is therefore 
highly desirable to accelerate the approach to the continuum
by cancelling at least the leading cutoff effects proportional to $a$.
The theoretical framework for this goes under the name of Symanzik
improvement~\cite{SymanzikI,SymanzikII}, and without any loss it may
be restricted to on-shell quantities and correlation functions
at physical distances~\cite{OnShell}.

The structure of the on-shell O($a$) improved lattice action 
has been derived a long time ago by Sheikholeslami and
Wohlert~\cite{SW}. A detailed discussion of the on-shell improved
theory with $\Nf$ mass degenerate quarks can be found in 
refs.~\cite{letter,paperI}. There, it was also shown how chiral symmetry
may be used to determine some of the improvement coefficients
non-perturbatively. In particular, the improved action
has been determined for both $\Nf=0$ (quenched approximation) and
$\Nf=2$~\cite{paperIII,KarlRainer}. Concerning the 
improvement of composite fields, non-perturbative results have been obtained 
for quark bilinear operators without derivatives. So far all results
are for the quenched theory. However, while the
methods used to determine the coefficients $\ca$~\cite{letter,paperIII}, 
$\cv$~\cite{MarcoRainer} and $c_{\rm T}$~\cite{Gupta1,Gupta2}  
also apply for $\Nf\neq 0$, the situation
is less clear for the $b$-coefficients, which multiply cutoff
effects proportional to the subtracted bare quark mass $\mq=m_0-\mc$. 

In the quenched approximation all the $b$-coefficients can be determined
from chiral Ward identities, by considering 
the more general case of mass non-degenerate
quarks~\cite{GiuliaRoberto,Gupta1,Gupta2}. Unfortunately these methods
do not easily generalise to the full theory, as the theory with
non-degenerate quarks requires many more O($a$) 
counterterms~(for the case $\Nf=3$ cp.~\cite{Sharpe99}).
Non-perturbative quenched results are known for $\bv$~\cite{paperIV}, 
for $\bm$  and for the combination $\ba-\bp$~\cite{GiuliaRoberto}. 
Furthermore, results for all coefficients have 
been published in refs.~\cite{Gupta2,Gupta3}. Finally we mention 
that all relevant improvement coefficients are 
known to one-loop order of perturbation 
theory~\cite{Wohlert-PeterStefan}

In the present paper we restrict ourselves to the determination of
$\bm$ and the combination $\ba-\bp$ in the quenched approximation. 
These coefficients are needed for the determination of renormalized quark
masses using either the bare subtracted or the bare PCAC quark masses
as starting point~\cite{strange1,strange2}. Motivated by this application
we thus  extend the previous study of ref.~\cite{GiuliaRoberto} 
to larger bare couplings, and also investigate the O($a$) 
ambiguities in these coefficients.

The paper is organized as follows. In sect.~2 we recall 
the PCAC relation and its  generalization to the
theory with non-degenerate quarks in the quenched approximation. 
It is then shown how the improvement coefficients can be isolated by
defining appropriate ratios of current quark masses.
Evaluating these both in perturbation theory and non-perturbatively
motivates our strategy (sect.~3). The results, 
together with some checks and technical details
are given in sect.~4, and we conclude with a summary of our findings.


\section{The PCAC relation}

In this section we assume that the reader is familiar with
ref.~\cite{paperI}. In particular we refer to this reference
for unexplained conventions and notation.

\subsection{Mass degenerate quarks}

Our starting point is the PCAC relation
\begin{equation}
   \partial_\mu A_\mu^a = 2m P^a,
\end{equation}
where $\psi$ is a doublet of light quarks and
\begin{equation}
  A_\mu^a = \psibar \gamma_\mu\gamma_5\frac{\tau^a}{2}\psi,\qquad
  P^a = \psibar \gamma_5 \frac{\tau^a}{2}\psi,
\end{equation}
are the isovector axial current and density respectively.
On the lattice with Wilson quarks this relation only
holds up to lattice artifacts of O($a$). These may be reduced
to O($a^2$) by tuning the coefficient $\csw$ of the 
Sheikholeslami-Wohlert term in the action, and the
coefficient $\ca$ in the improved axial current,
\begin{equation}
  (A_I)_\mu^a = A_\mu^a + \ca a\tilde\partial_\mu P^a,
\end{equation}
such that the PCAC relation with the improved current
holds exactly in a few selected matrix elements. 
In this way, non-perturbative results 
have been obtained for $\ca$ and $\csw$ in the quenched
approximation~\cite{paperIII} and for $\csw$ also in the case 
$\Nf=2$~\cite{KarlRainer}. 

Given the improved action and the improved axial current,
the PCAC relation leads to the definition of a renormalized
O($a$) improved quark mass,
\begin{equation}
  \mr = {{\Za(1+\ba a\mq)}\over{\Zp(1+\bp a\mq)}} m.
\end{equation}
Here, $m$ is a bare current quark mass defined through some
matrix element of the PCAC relation.
On the other hand, the renormalized quark can be written in the form
\begin{equation}
   \mr = \Zm\mqtilde,\qquad  \mqtilde=\mq(1+\bm a\mq),
\end{equation}
where $\mq=m_0-\mc$ is the subtracted bare quark mass.
Combining these equations and expanding in powers of $a$ we obtain 
\begin{equation}
  m = Z\,\mq\Bigl(1+ \left[\bm-\ba+\bp\right] a\mq\Bigr) +\rmO(a^2),
 \label{PCACmass}
\end{equation}
where $Z$ is a finite combination of renormalization constants,
\begin{equation}
  Z={\Zm\Zp\over\Za},
\end{equation}
which is a function of $\gtilde^2 = g_0^2(1+\bg a\mq)$ only. 
From this equation it is clear that the combination $\bm-\ba+\bp$ 
of improvement coefficients can be determined by 
studying the dependence of $m$ upon $\mq$ at fixed $\gtilde$. 
If instead $g_0$ is kept constant, one obtains an additional contribution,
\begin{equation}
   m = Z\mq\left(1+ \left[\bm-\ba+\bp +
       g_0^2\, {\partial\ln Z \over\partial g_0^2}\,\bg \right]a\mq\right) 
      +\rmO(a^2),
\end{equation}
where the renormalization constant is now evaluated at $g_0^2$.
We note however, that knowledge of the combination $\bm-\ba+\bp$ 
is not sufficient to determine a renormalized improved quark mass, 
if one starts from the (measurable) subtracted bare quark mass $\mq=m_0-\mc$
or the bare PCAC mass $m$. Beside the determination of the
renormalization constants it is then necessary to determine
$\bm$ and the combination $\ba-\bp$ separately. 



\subsection{Non-degenerate quarks and the quenched approximation}

It has been noticed in ref.~\cite{GiuliaRoberto} and later in
ref.~\cite{Gupta1} that the quenched (or valence) 
approximation leads to major simplifications, which may
be used to determine the coefficients $\ba,\bp$ and $\bm$ separately. 
First of all the coefficient $\bg$ vanishes in this approximation. 
But more importantly, the structure of the O($a$) improved theory 
with non-degenerate quarks remains relatively simple, in contrast
to the full theory, where non-degenerate quarks lead to a proliferation
of new improvement coefficients~\cite{Sharpe99}. 
In the remainder of the paper we will restrict ourselves to the
quenched approximation and shortly review the features that
will be needed in the sequel.

In the quenched approximation the bare quark masses are 
separately improved for each quark flavour, i.e. an equation of the form
\begin{equation}
  \mqtilde = \mq(1+\bm a\mq),
\end{equation}
holds for each flavour. The isospin symmetry is broken 
in the presence of non-degenerate masses, and it is useful 
to introduce the off-diagonal bilinear fields through
\begin{equation}
  A_\mu^{\pm} = A_\mu^1\pm iA_\mu^2,\qquad P^{\pm} = P^1\pm iP^2,
\end{equation}
which are expected to satisfy the PCAC relation
\begin{equation}
    \partial_\mu A_\mu^{\pm} = (m_{\rm u} + m_{\rm d})P^{\pm},
\end{equation} 
up to cutoff effects\footnote{The flavours in the
isospin doublet are generically denoted by $u$ and $d$. Of course, 
one may also identify one of the flavours with the strange quark.}.
In the quenched approximation the improvement of these off-diagonal
fields is the same as in the degenerate case, except that
$\ba$ and $\bp$ now multiply the average of the subtracted bare
quark masses. For instance the improved axial current is given by
\begin{equation}
  (A_{\sixrm R})_\mu^{\pm} = \Za\left[1+\ba\frac12
            \left(am_{\rm q,u}+am_{\rm q,d}\right)\right](A_I)_\mu^\pm,
\end{equation}
and the axial density has an analogous structure. 
The mass dependence is such that
$\bm$ can be disentangled from $\ba-\bp$ by considering the
analogue of eq.~(\ref{PCACmass}) for the case of non-degenerate quarks.

\subsection{Current quark masses and estimators for $\bm$, $\ba-\bp$ and $Z$}

To define current quark masses we
use correlation functions derived from the QCD Schr\"odinger
functional~\cite{LNWW,StefanI,paperI},
\begin{equation}
  \fa^{ij}(x_0)= -\frac12\langle A_0^{+}(x)O^{-}\rangle,\qquad
  \fp^{ij}(x_0)= -\frac12\langle P^{+}(x)O^{-}\rangle,
\end{equation}
with the source $O^{\pm}=O^1\pm iO^2$ and
\begin{equation}
   O^a = a^6\sum_{\bfy,\bfz}\zetabar(\bfy)\gamma_5\frac{\tau^a}{2}\zeta(\bfz).
\end{equation}
The indices $i,j$ refer to choices $m_{0,{\rm u}}=m_{0,i}$ and 
$m_{0,{\rm d}}=m_{0,j}$ for the quark masses 
taken from a list of numerical values $ \{m_{0,1}, m_{0,2}, \ldots \}$
to be specified later.
This definition
is such that the standard correlation functions $\fa$ and
$\fp$~\cite{letter,paperI} are recovered in the degenerate case $i=j$.
We now define bare current quark masses through
\begin{equation}
   m_{ij} = {\tilde\partial_0\fa^{ij}(x_0)
          +a\ca\partial_0^\ast\partial_0^{}\fp^{ij}(x_0)\over{2\fp^{ij}(x_0)}}.
  \label{current_mass}
\end{equation}
Apart from $x_0$ these quark masses depend on the lattice 
size $L/a$, the ratio $T/L$, the angle $\theta$~\cite{paperI} 
and also the precise definition of the derivatives. 
In eq.~(\ref{current_mass}) 
we have followed ref.~\cite{paperIII} by
setting $\tilde\partial_0=\frac12(\partial_0+\partial_0^\ast)$,
with the usual forward and backward derivatives given by
\begin{eqnarray}
  \partial_0      f(t) &=& {1\over a} \left[f(t+a) - f(t)\right],\\ 
  \partial_0^\ast f(t) &=& {1\over a} \left[f(t) - f(t-a)\right].
\end{eqnarray}
In addition we consider current quark masses 
with improved derivatives~\cite{GiuliaRoberto}, 
by replacing in eq.~(\ref{current_mass}),
\begin{eqnarray}
  \tilde\partial_0 &\rightarrow & 
  \tilde\partial_0\left(1-\frac16a^2\partial_0^\ast\partial_0^{}\right),\\
  \partial_0^\ast\partial_0^{}&\rightarrow &
  \partial_0^\ast\partial_0^{}\left(1-\frac1{12}a^2
  \partial_0^\ast\partial_0^{}\right).
\end{eqnarray}
When acting on smooth functions these improved lattice derivatives
have errors of O($a^4$) only.
The generalization of eq.~(\ref{PCACmass}) now reads
\begin{eqnarray}
   m_{ij} &=& Z\Bigl[\frac12(m_{{\rm q},i}+m_{{\rm q},j})
            +\frac12\bm(am_{{\rm q},i}^2+ am_{{\rm q},j}^2)\nonumber\\
           &&\hphantom{012345}
           -\frac14 (\ba-\bp)a(m_{{\rm q},i}+m_{{\rm q},j})^2\Bigr]
            +\rmO(a^2).
\end{eqnarray}
To isolate the coefficients we consider the combination
\begin{equation}
  2am_{12}-am_{11}-am_{22}=f(am_{{\rm q},1},am_{{\rm q},2}),
\end{equation}
which is an analytic function of the subtracted bare quark masses. 
Furthermore, it is symmetric under exchange of the arguments,
$f(x,y)=f(y,x)$, and vanishes for $x=y$, so that
its series expansion can be cast in the form
\begin{equation}
  f(x,y) = (x-y)^2\sum_{n,k=0}^{\infty}c_{nk}(x-y)^{2n}(x+y)^k,
\end{equation}
with real coefficients $c_{nk}$. In particular, we note that~\cite{GiuliaRoberto}
\begin{equation}
  c_{00}= Z\frac12(\ba-\bp),
\end{equation}
up  to terms of O($a$) which do not depend 
on the quark masses. To isolate $\ba-\bp$ 
we also consider the difference
\begin{equation}
   am_{11}-am_{22} =  g(am_{{\rm q},1},am_{{\rm q},2}),   
\end{equation}
and obtain an expansion of the form
\begin{equation}
  g(x,y) = (x-y)\sum_{n,k=0}^{\infty}d_{nk}(x-y)^{2n}(x+y)^k,
\end{equation}
with $d_{00} = Z$ (up to quark mass independent terms of
order $a^2$). Hence it is clear that the ratio
\begin{equation}
  \Rap = {{2(2m_{12}-m_{11}-m_{22})}\over{(m_{11}-m_{22})
         (am_{{\rm q},1}-am_{{\rm q},2})}},    
  \label{R_AP}
\end{equation}
has a chiral limit and provides an estimate for the coefficient
$\ba-\bp$, up to terms of O($am_{{\rm q},1}+am_{{\rm q},2}$)
and other quark mass independent lattice artifacts of O($a$).

To obtain a similar estimate for $\bm$ it is useful to introduce 
a third quark mass given by the mean of the first two,
\begin{equation}
  \label{eq:m3}
  m_{0,3}=\frac12(m_{0,1}+m_{0,2}).
\end{equation}
Taking the same steps as above we can estimate $\bm$ through
\begin{equation}
  R_{\rm m}=   
  {{4(m_{12}-m_{33})}\over{(m_{11}-m_{22})(am_{{\rm q},1}-am_{{\rm q},2})}},   
 \label{R_m} 
\end{equation}
where the error is again O($am_{{\rm q},1}+am_{{\rm q},2}$).
We note in passing that the combination $\ba-\bp-\bm$
can be estimated from the ratio 
\begin{equation}
   \Rap-\Rm = {{2(2m_{33}-m_{11}-m_{22})}\over{(m_{11}-m_{22})
         (am_{{\rm q},1}-am_{{\rm q},2})}}, 
  \label{Rapm}
\end{equation}
which involves only correlation functions with mass degenerate quarks
(cf. subsect.~2.1).
Furthermore one could estimate the finite renormalization constant
$Z$ through
\begin{equation}
 R_Z = {{m_{11}-m_{22}}\over{m_{{\rm q},1}-m_{{\rm q},2}}}
         + (\ba-\bp-\bm)(am_{11}+am_{22}),
 \label{R_Z}
\end{equation}
which is correct up to O($a^2$) for the correct choice of 
$\ba-\bp-\bm$. However, knowledge of this combination is not required as
one may instead use the estimate~(\ref{Rapm}). 
Alternative estimates of $Z$ can be obtained from current quark masses 
which derive from correlation functions with  non-degenerate quarks. 
Also in this case it is possible to cancel errors of O($am$) 
by using the estimates $\Rap$ and $\Rm$. 
Finally we emphasize that all the ratios 
have a chiral limit, and none of them 
requires knowledge of the critical mass $\mc$.
\begin{table}[htb]
\centering
\begin{tabular} {|c|ll|ll|ll|}
\hline
  &\multicolumn{1}{c}{}&\multicolumn{1}{c|}{}&\multicolumn{1}{c}{}
  &\multicolumn{1}{c|}{}&\multicolumn{1}{c}{}&\multicolumn{1}{c|}{}\\[-2ex]
 $L/a$ &\multicolumn{1}{c}{$\Rap^{(0)}$}&\multicolumn{1}{c|}{$\Rap^{(1)}$} 
       &\multicolumn{1}{c}{$R_{\rm m}^{(0)}$}&\multicolumn{1}{c|}
           {$R_{\rm m}^{(1)}$} 
       &\multicolumn{1}{c}{$R_Z^{(0)}$}&\multicolumn{1}{c|}{$R_Z^{(1)}$} \\[1ex]
\hline 
\multicolumn{7}{|c|}{}\\[-2ex]
\multicolumn{7}{|c|}{standard derivatives}\\[.5ex]
\hline
8 &$-0.0591$& $-0.0544$& $-0.4628$& $-0.0819$ & 1.0022& 0.0881\\
12&$-0.0405$& $-0.0381$& $-0.4753$& $-0.0867$ & 1.0010& 0.0895\\
16&$-0.0307$& $-0.0297$& $-0.4815$& $-0.0891$ & 1.0005& 0.0900\\
20&$-0.0248$& $-0.0245$& $-0.4853$& $-0.0905$ & 1.0003& 0.0902\\
24&$-0.0207$& $-0.0210$& $-0.4877$& $-0.0915$ & 1.0002& 0.0903\\
\hline
  \multicolumn{7}{|c|}{}\\[-2ex]
  \multicolumn{7}{|c|}{improved derivatives}\\[.5ex]
\hline 
8 & \hphantom{+}0.0065& \hphantom{+}0.0035& $-0.4734$& $-0.0850$& 0.9973& 0.0856\\
12& \hphantom{+}0.0039& \hphantom{+}0.0024& $-0.4824$& $-0.0890$& 0.9988& 0.0885\\
16& \hphantom{+}0.0028& \hphantom{+}0.0018& $-0.4869$& $-0.0910$& 0.9993& 0.0894\\
20& \hphantom{+}0.0022& \hphantom{+}0.0014& $-0.4896$& $-0.0921$& 0.9996& 0.0898\\ 
24& \hphantom{+}0.0018& \hphantom{+}0.0011& $-0.4913$& $-0.0929$& 0.9997& 0.0900\\
\hline\hline
$\infty$& \hphantom{+}0.0& $-0.0009$ &$-0.5$& $-0.0962$& 1.0& 0.0905\\
\hline 
\end{tabular}
\caption[Table]{\footnotesize 
The estimates for the improvement coefficients and the renormalization
constant in perturbation theory on lattices of size $T/a\times(L/a)^3$
with $T=3L/2$. We have set $x_0=T/2$, $\theta=0.5$ and chosen the bare
quark masses such that $Lm_{\rm q,1}=0.24$ and $Lm_{\rm q,2}=0.40$
for all lattice sizes. The expected limiting values for
the coefficients are given in the last row. 
Corrections at finite lattice size are O($a/L$) for the improvement
coefficients and O($a^2/L^2$) for the renormalization constant (up to
logarithms).}
\label{ex_PT}
\end{table}

\subsection{Perturbation theory}

The perturbative expansion of $\fa$ and $\fp$ on a lattice
of fixed size $L/a$ has been 
explained in ref.~\cite{paperII,PeterStefan}. 
It is then straightforward to obtain the corresponding expansion of
the current quark masses and hence of the ratios $R_{\rm X}$ 
(${\rm X}= {\rm AP,m},Z$), viz.
\begin{equation}
   R_{\rm X}= R_{\rm X}^{(0)}+g_0^2 R_{\rm X}^{(1)}+\rmO(g_0^4).
\end{equation}
After extrapolation to the chiral limit the coefficients
are still affected by quark mass independent lattice artifacts.
However, as $L$ is then the only scale in the 
problem, these effects may be cancelled by extrapolating
the lattice size $L/a$ to infinity. Alternatively one may
fix the subtracted bare quark masses in units of $L$, and then take 
the limit $L/a \rightarrow\infty$, i.e.~the
chiral and the infinite volume limit are reached at the same time.
An example is given in table~1.
As expected the estimates for the coefficients converge to the 
known results of ref.~\cite{PeterStefan}, 
which were obtained by requiring scaling of 
renormalized finite volume correlation functions.
For gauge group SU(3) and neglecting terms of O($g_0^4$) 
one has
\begin{eqnarray}
    \bm      &=& -\frac12 -0.09623(3)\times g_0^2, \label{bm_pert}\\[1ex] 
    \ba-\bp  &=& -0.00093(8)\times g_0^2.          \label{babp_pert}
\end{eqnarray}
The perturbative result for the 
renormalization constant $Z$ is implicit in the
literature~\cite{GabrielliEtAl} and has been computed by one of
the authors~\cite{Stefan_notes}. For $N=3$ it is given by
\begin{equation}
   Z= 1+ 0.090514(2) \times g_0^2   + \rmO(g_0^4).  \label{Z_pert}
\end{equation}

\section{Improvement conditions}

\subsection{Improvement conditions at fixed lattice size}

Beyond perturbation theory the estimates for the 
improvement coefficients are affected by an intrinsic
ambiguity of order $a/r_0$ ($r_0$ denotes the hadronic scale
of ref.~\cite{Rainer_r0}). In contrast to the perturbative
example of the last section it is thus impossible
to eliminate this ambiguity by chiral and infinite volume extrapolations. 
Therefore the estimates $R_{\rm X}$ at some fixed lattice 
size $L/a$, at fixed quark masses $am_{0,i}$ 
and for a definite choice of $T/L$ and
$\theta$ may be taken as a {\em definition} of the improvement
coefficients, provided the O($a\mq$) and O($a/L$) lattice artifacts 
are not too large. At our strongest coupling $\beta=6/g_0^2=6.0$
we expect typical O($a/r_0$) ambiguities to be around 0.1 or less.
Taking this as a guideline we use perturbation theory 
to choose the lattice size and the quark mass parameters.

A typical example is summarized in table~2. The lattice
size is $12\times 8^3$, the angle $\theta=0.5$, and
the bare masses are $am_{\rm q,1}=0.03$ and $am_{\rm q,2}=0.05$.
To obtain the one-loop coefficients we have used $\ca^{(1)}=-0.00757$ 
and the critical quark mass has been set to its value in infinite
volume~\cite{paperII}. However, as expected, the dependence upon the latter
is weak.
\begin{table}[htb]
\centering
\begin{tabular} {|c|rr|rr|rr|}
\hline
  &\multicolumn{1}{c}{}&\multicolumn{1}{c|}{}&\multicolumn{1}{c}{}
  &\multicolumn{1}{c|}{} 
  &\multicolumn{1}{c}{}&\multicolumn{1}{c|}{}\\[-2ex]
 $x_0$ &\multicolumn{1}{c}{$\Rap^{(0)}$}&\multicolumn{1}{c|}{$\Rap^{(1)}$} 
       &\multicolumn{1}{c}{$R_{\rm m}^{(0)}$}&\multicolumn{1}{c|}
           {$R_{\rm m}^{(1)}$} 
       &\multicolumn{1}{c}{$R_Z^{(0)}$}&\multicolumn{1}{c|}{$R_Z^{(1)}$} \\[1ex]
\hline 
\multicolumn{7}{|c|}{}\\[-2ex]
\multicolumn{7}{|c|}{standard derivatives}\\[.5ex]
\hline
3& $-0.052$& $-0.035$& $-0.456$& $-0.086$ & 1.002& 0.089\\
4& $-0.054$& $-0.039$& $-0.458$& $-0.082$ & 1.002& 0.090\\
5& $-0.056$& $-0.046$& $-0.460$& $-0.081$ & 1.002& 0.089\\
6& $-0.059$& $-0.054$& $-0.463$& $-0.082$ & 1.002& 0.088\\
7& $-0.062$& $-0.062$& $-0.466$& $-0.084$ & 1.002& 0.087\\
8& $-0.065$& $-0.068$& $-0.469$& $-0.085$ & 1.002& 0.085\\
9& $-0.069$& $-0.069$& $-0.472$& $-0.086$ & 1.002& 0.084\\
\hline
  \multicolumn{7}{|c|}{}\\[-2ex]
  \multicolumn{7}{|c|}{improved derivatives}\\[.5ex]
\hline 
3& 0.009& 0.020& $-0.471$& $-0.098$& 0.997& 0.094\\
4& 0.008& 0.005& $-0.472$& $-0.087$& 0.997& 0.088\\
5& 0.007& 0.004& $-0.473$& $-0.086$& 0.997& 0.086\\
6& 0.007& 0.004& $-0.473$& $-0.085$& 0.997& 0.086\\
7& 0.006& 0.003& $-0.474$& $-0.085$& 0.997& 0.085\\
8& 0.005& 0.002& $-0.475$& $-0.085$& 0.997& 0.085\\
9& 0.003& 0.011& $-0.477$& $-0.086$& 0.997& 0.089\\
\hline 
\end{tabular}
\caption[Table]{\footnotesize 
The estimates for the improvement coefficients and the renormalization
constant $Z$ in perturbation theory on a $12\times8^3$ lattice.
The parameters are as explained in the text. The 
one-loop coefficients have been evaluated for gauge group SU(3).}
\label{ex_PT1}
\end{table}

%
We note that the improved derivative
yields consistently better results also at the one-loop level,
where this is a priori not expected. 
With the chosen parameters, the perturbative results show the expected
behaviour for the cutoff effects. In particular, we have checked that
the  deviation from the exact results becomes smaller with decreasing
quark masses. 

\begin{figure}[h]
\begin{center}
\epsfig{figure=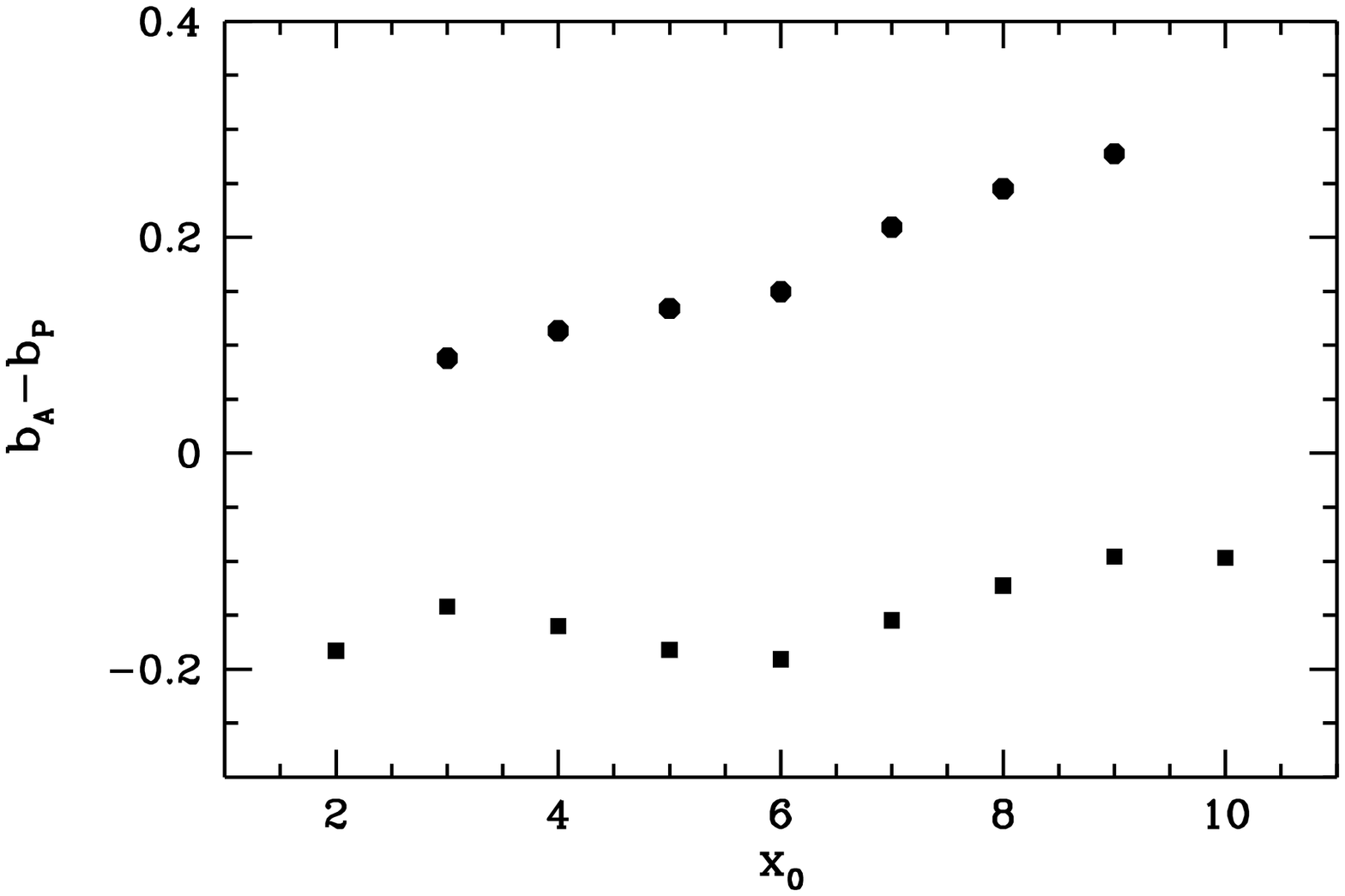, width=12.5 true cm}
\end{center} 
\vskip -1.0 true cm
\caption{\footnotesize
Estimates of $\ba-\bp$ versus $x_0$. Squares are obtained with the 
standard, circles with the improved derivative.
\label{babp_spread}}
\end{figure}

\begin{figure}[h]
\begin{center}
\epsfig{figure=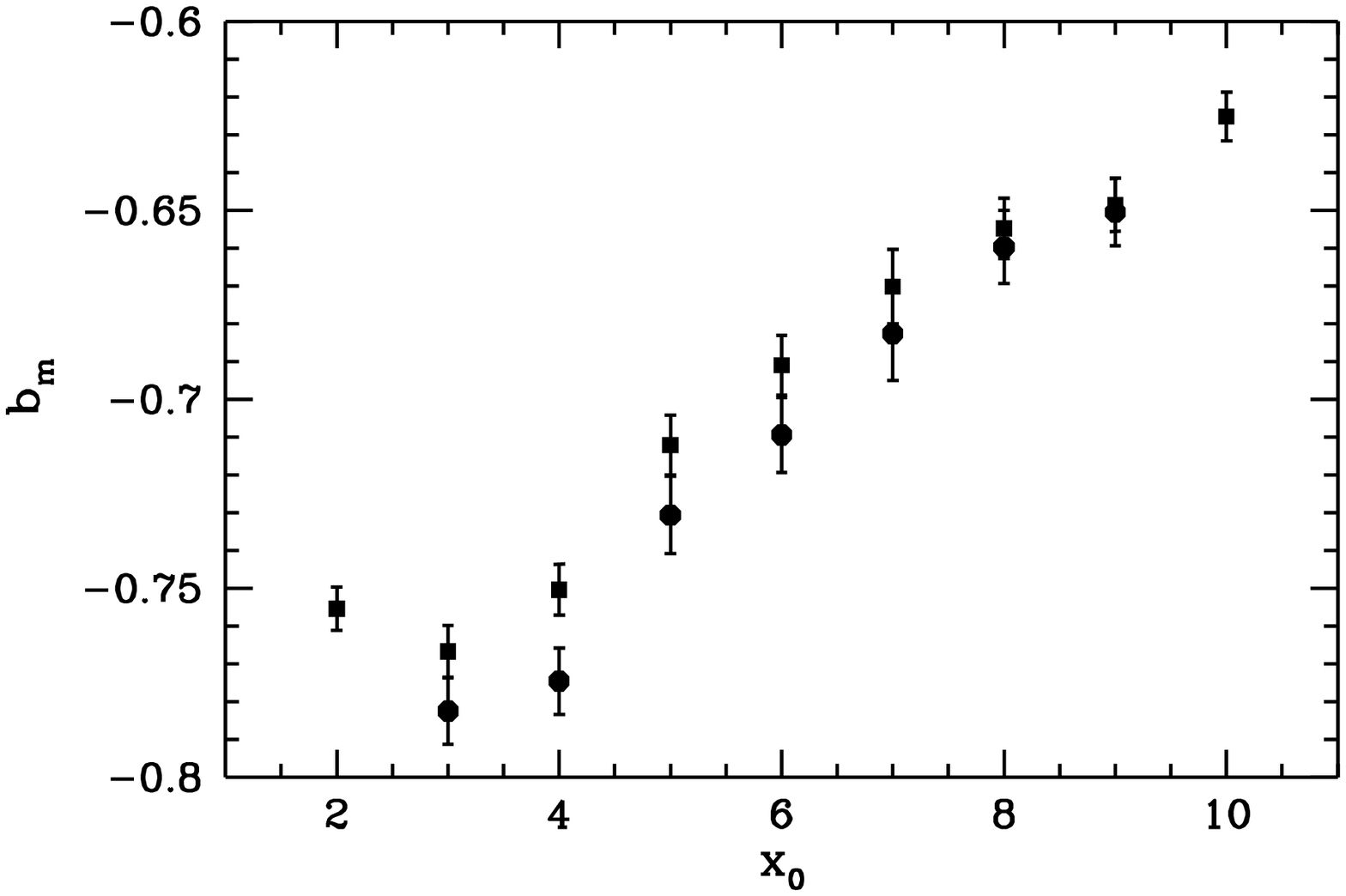, width=12.5 true cm}
\end{center} 
\vskip -1.0 true cm
\caption{\footnotesize
Estimates of $\bm$ versus $x_0$. Squares are obtained with the 
standard, circles with the improved derivative.
\label{bm_spread}}
\end{figure}

Evaluating the same estimators non-perturbatively at $\beta=6.0$,
with approximately the same choice of parameters
(cp.~sect.~4), and with the non-perturbative 
values for $\csw$ and $\ca$~\cite{paperIII}, we obtain the results
shown in figs.~\ref{babp_spread} and \ref{bm_spread}.
Somewhat surprisingly the non-perturbative results for $\ba-\bp$
are quite sensitive to the choice of parameters, 
and to the choice of the lattice derivative. The situation may be 
slightly improved by correcting for the tree-level
cutoff effects, i.e.~by setting 
\begin{equation}
   \ba - \bp = \Rap -\Rap^{(0)}, \qquad
           \bm  = \Rm  -  \Rm^{(0)} - \frac12,
\end{equation}
such that the estimators assume the correct values
in the continuum limit.
However, from the perturbative results of table~2 we infer
that this only corrects for a fraction of the discrepancy
in the case of $\ba-\bp$, while the $x_0$ dependence of
the estimator for $\bm$ becomes somewhat more pronounced.

\subsection{Improvement conditions at constant physics}

At least the case of $\ba-\bp$  seems different from the
determination of $\csw$ and $\ca$ in ref.~\cite{paperIII}, where the 
results were quite stable against a variation of 
the parameters. 
Already in perturbation theory the ${\rm O}(a)$ effects in $\ba-\bp$
are much larger than in $\csw$ and $\ca$.
One might argue that our quark masses are not small
enough, or that the lattice size is too small. However, 
it is our general experience that the spread of 
the estimates is considerably larger than naively expected, 
also on larger lattices or if a chiral extrapolation is attempted.
It should be emphasized at this point that this does not imply a failure of
the improvement programme. Improvement is an 
asymptotic concept, and thus only determines the {\em rate} 
of the continuum approach. It is a priori not clear how large 
the intrinsic O($a$) ambiguities typically are for a 
given improvement coefficient.
 
There are essentially two alternative ways of dealing with this situation.
First one may determine a ``typical'' spread of the values for $R_{\rm X}$
(${\rm X}\neq Z$) and  quote a corresponding systematic error for the estimate 
of the improvement coefficient. 
In this paper we follow a second approach 
by imposing the improvement condition at constant physics. 
More precisely, we fix all renormalized quantities in units of $L$ 
and keep $L/r_0$ constant as we approach the continuum limit.
As a consequence the estimates become
smooth functions of $a/r_0$ and thus also of $g_0$.
Furthermore it is obvious that any other estimate 
$\tilde{R}_{\rm X}$ may have a different dependence upon $g_0$,
but the difference $ R_{\rm X}- \tilde{R}_{\rm X}$
is again a smooth function which must vanish in the continuum 
limit with a rate proportional to $a/r_0$.
Note that the very same  strategy has previously been applied
to the determination of the finite renormalization constants for
the isospin currents~\cite{paperIV}. However, in the case
of renormalization constants the intrinsic ambiguity is of O($a^2$).


Our procedure highlights the importance of the continuum limit. 
Furthermore we emphasize that in cases where the intrinsic O($a$) ambiguity
is large, the important result is not so much a numerical 
value for the improvement coefficient at fixed $\beta$ but
rather its dependence on $g_0$ which follows
from keeping the physics fixed.

\begin{figure}[h]
\begin{center}
\epsfig{figure=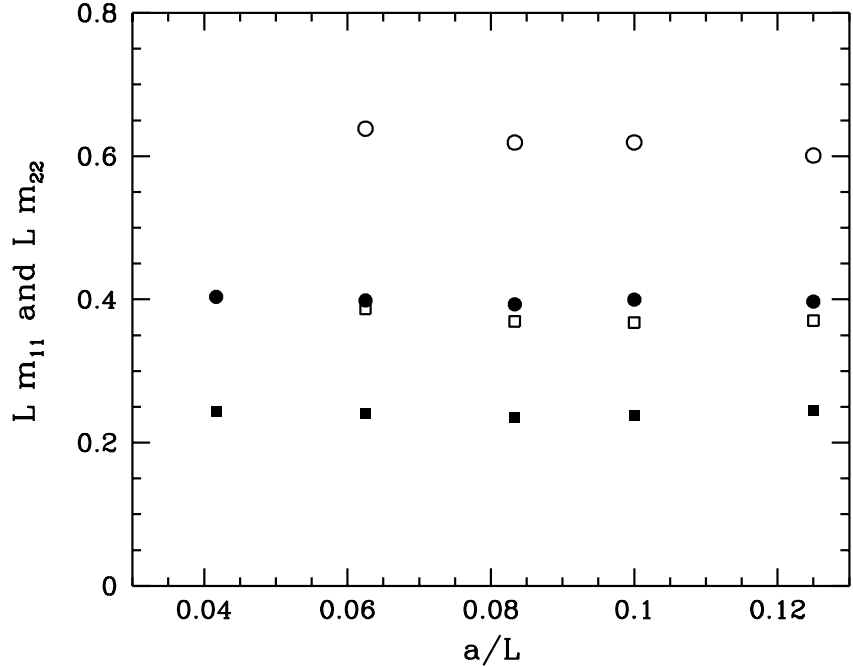, width=12.5 true cm}
\end{center} 
\vskip -1.0 true cm
\caption{\footnotesize
This plot illustrates how well the physics is
kept constant. Full symbols refer to the bare current quark masses,
empty symbols to the renormalized quark masses (the latter are not available
at the largest $\beta$ value).
\label{mass_tuning}}
\end{figure}

\section{Technical details and results}

We describe the technical aspects of the numerical 
simulations in detail and present our results.

\subsection{Choice of simulation parameters}

The improved action and axial current have been determined
for $\beta=6/g_0^2\geq 6.0$~\cite{paperIII}. This sets the upper
limit for the bare couplings to be considered.
At $\beta=6.0$ the probability of encountering
a quark zero mode (so-called ``exceptional configurations'')
increases rapidly if the lattice volume is too large
or the quark mass too small~\cite{paperIII}. 
To avoid this problem we choose a $12\times 8^3$ lattice, $\theta=0.5$ 
and quark masses such that $am_{11} =0.03$ and $am_{22}=0.05$.
The parameters are thus approximately the same as
in the perturbative example discussed in the preceding section.
Note that by using the current quark masses instead of 
the subtracted bare quark masses we avoid the determination of 
the critical quark mass. In fact, as none of our observables
requires the knowledge of $\mc$ we will not determine
this parameter at all.

We want to keep the physics constant 
as we vary $\beta$ towards larger values. To this end we fix
all dimensionfull quantities in units of the
physical scale $r_0$~\cite{Rainer_r0},
which has been determined very precisely in ref.~\cite{r0_ref}.
From the fit function,
\begin{eqnarray}
  \ln(a/r_0) &=& -1.6805 - 1.7139 (\beta-6) + 0.8155 (\beta-6)^2\nonumber\\
             && \hphantom{0123}- 0.6667 (\beta-6)^3, 
  \label{r0fit}
\end{eqnarray}
which is valid in the interval $5.7\leq \beta \leq 6.57$,
we infer that our lattice size $L/a=8$ at $\beta=6.0$ 
corresponds to $L/r_0=1.49$. The requirement that this ratio 
be constant as the lattice size $L/a$ is increased determines the
corresponding $\beta$ values. Furthermore, we choose $T=3L/2$ 
and tune the  bare quark masses $m_{0,1}$ and $m_{0,2}$ 
such as to maintain
\begin{equation}
  Lm_{11} = 0.24,\qquad  Lm_{22}=0.40 
  \label{mL}
\end{equation}
for all lattice sizes with the third bare quark mass then given
by (\ref{eq:m3}).
The resulting simulation parameters are summarized in table~\ref{tab:n1}. 
Note that we quote the values for the hopping parameter $\kappa=1/(2am_0+8)$
rather than $am_0$, since we use this parameter in our programs.
A complication arises at our largest lattice size, $L/a=24$,
as the corresponding $\beta$ value lies clearly outside 
the range of validity of eq.~(\ref{r0fit}). 
In order to include the largest lattice we have 
extrapolated the data points in~\cite{r0_ref} using a linear fit 
of $\ln(a/r_0)$ as a function of $\beta$. 
Of course this procedure is not unique. 
From the spread of results obtained with different 
ans\"atze for the fit we estimate the error in $\beta$ to be at most
$0.003$. However, from this procedure it is not possible to estimate
the real error in a reliable way
and the results obtained at $\beta=6.756$ have to be 
taken with some care.
%
\begin{table}[htbp]
\begin{center}
  \begin{tabular}{|l|l|l|l|} \hline
    $L/a$ & $\beta$      & $\kappa_1$   & $\kappa_2$\\ \hline
    8     & $6.0$        & $0.133931$   & $0.133241$\\
    10    & $6.13826$    & $0.134828$   & $0.134263$\\
    12    & $6.26229$    & $0.135107$   & $0.134654$\\
    16    & $6.46895$    & $0.135126$   & $0.134794$\\
    24    & $6.756$      & $0.134856$   & $0.134634$\\ \hline
  \end{tabular}
  \caption{\footnotesize
    Values for $\beta$, $\kappa_1$ and $\kappa_2$ for simulations
    with constant physical volume and with $Lm_{11}=0.24$ and $Lm_{22}=0.4$.}
  \label{tab:n1}
\end{center}
\end{table}
In fig.~\ref{mass_tuning} one can see how precisely the conditions~(\ref{mL})
are satisfied. The relative accuracy is better than 2 per cent. 
The reader may be worried that we are keeping
fixed the bare current quark mass in units of $L$, rather than 
corresponding renormalized masses. It turns out that
the difference is small in the range of $\beta$ values
considered. To illustrate this, we have, for $L/a\leq 16$, also plotted 
$L\mr = Lm\Za/\Zp$ for both masses. They are  obtained using the 
non-perturbative result for $\Za$~\cite{paperIV} and $\Zp$ in the SF
scheme~\cite{PeterStefanI,strange1}. The largest lattice had
to be left out, as a non-perturbative result for
$\Zp$ is not available for $\beta = 6.756$.
As the renormalization constant barely varies over the range 
of couplings considered, also the renormalized 
quark masses are constant to a good precision.

\subsection{The numerical simulation}

Our numerical simulations were performed on APE/Quadrics parallel
computers with SIMD architecture and single precision arithmetic following the
IEEE standard. We have used machines with up to 512 processors with an
approximate peak performance of 50 MFlops per node. 
While we have distributed the larger lattices over the whole machine the 
smaller lattices have been simulated with up to 32 independent replica.
To generate the
gauge fields we have applied a hybrid overrelaxation 
algorithm with two overrelaxation
steps per heatbath sweep. 
Every 15th complete update step the
fermionic correlation functions have been measured 
for all combinations of quark masses. 
With these choices 
the integrated autocorrelation time between successive measurements of
the fermionic correlations 
was found to be negligible. 
All in all we have accumulated between 870 and 3712 independent 
measurements for the various lattice sizes. 
From expectation
values of the fermionic correlation functions our secondary
observables~(\ref{R_AP}), (\ref{R_m}) and~(\ref{R_Z}) can be
constructed as described in sect.~2. 
To increase statistics we have chosen to average 
our secondary observables over the time slices $L/(2a),\ldots,(T-L/2)/a$.
Note that the number of time slices used for the average 
is scaled with the lattice size,
so that also in this respect the requirement of constant
physics is satisfied.
To compute the errors of $\ba-\bp$, $\bm$ and $Z$ 
we have calculated the autocorrelation
functions of these observables along the lines of appendix A
in~\cite{Ullis_appendix}.

\subsection{Result for $\ba-\bp$}
\label{sec:n1.2}

Our results are tabulated in the second column of table~\ref{tab:n2}
\begin{table}[t]
  \begin{center}
    \begin{tabular}{|l|l|l|l|l|}\hline
$\beta$   & $g_0^2 $   & $\ba-\bp$     & $\bm$             & $Z$         \\ \hline
$6.0$     & $1.0$      & $0.1703(49)$  & $-0.7127(48)(7)$  & $1.0605(4)$\\
$6.13826$ & $0.9775$   & $0.0563(23)$  & $-0.6915(35)(11)$ & $1.0907(4)$ \\
$6.26229$ & $0.9581$   & $0.0251(33)$  & $-0.6852(36)(30)$ & $1.0992(4)$ \\
$6.46895$ & $0.9275$   & $0.0090(52)$  & $-0.6846(44)(68)$ & $1.1048(3)$ \\
$6.756$   & $0.8881$   & $0.0007(36)$  &                   & $1.1047(2)$ \\ \hline
    \end{tabular}
    \caption{\footnotesize Non perturbative results for $\ba-\bp$, $\bm$ and $Z$.
      We have averaged over the time slices $L/(2a),\ldots,(T-L/2)/a$,
      and quote the corresponding statistical errors. In the case of $\bm$, 
      the second error is an estimate of the uncertainty 
      due to rounding errors in the determination of the third
      $\kappa$ value. Further details are given in the subsect.~4.4.}
    \label{tab:n2}
  \end{center}
\end{table}
\begin{figure}[htbp]
  \begin{center}
   \epsfig{figure=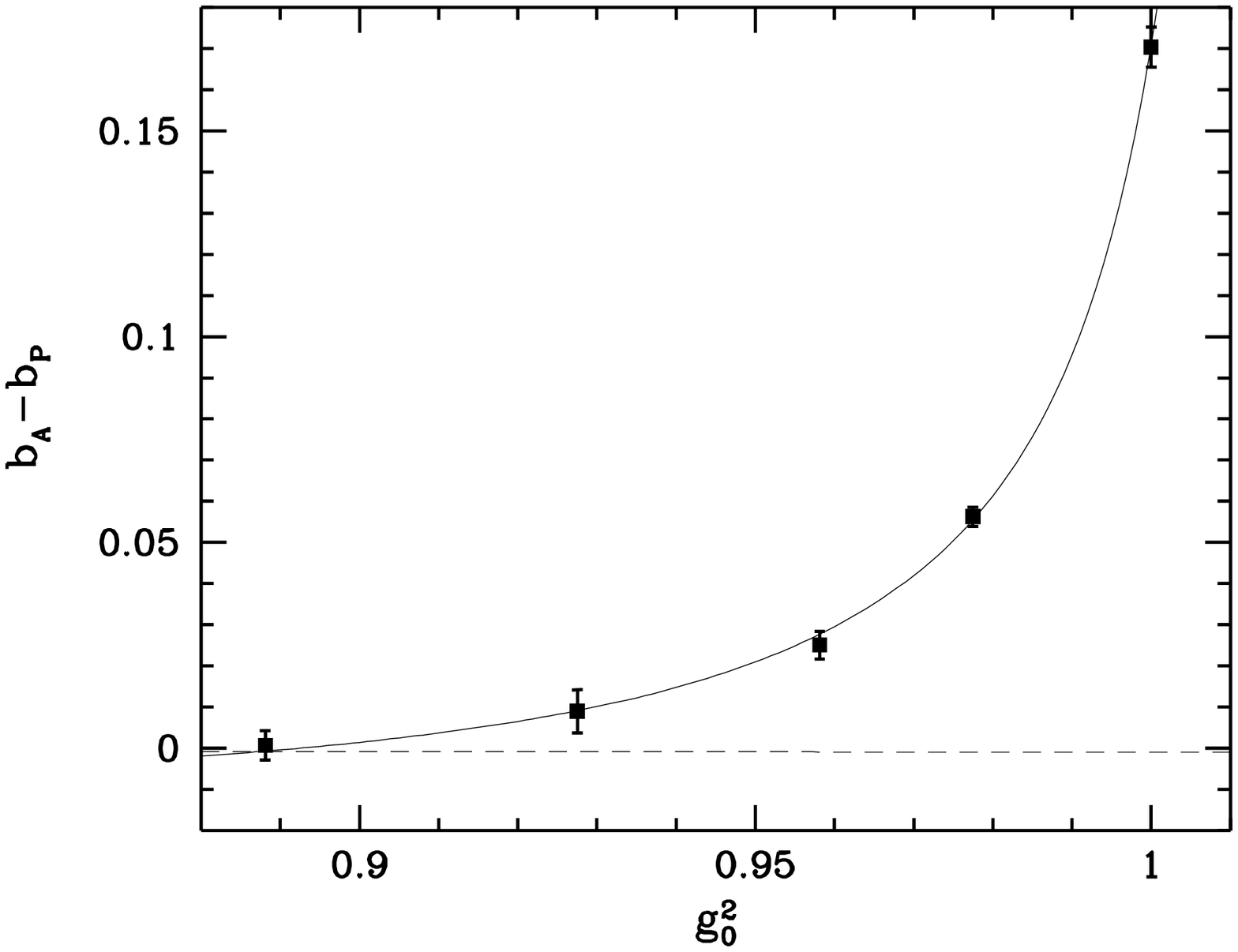, width=12.5cm}
    \caption{\footnotesize Results for $\ba-\bp$ as a function of $g_0^2$ from
      numerical simulations (filled squares) and one-loop perturbation
      theory (dashed line). The full line represents the
      fit~(\ref{babp_fit}).}
    \label{fig:n1}
  \end{center}
\end{figure}
and shown in figure~\ref{fig:n1}. At small values of the bare
coupling the non perturbative result is consistent with the one-loop
result~(\ref{babp_pert}). 
At larger couplings $\ba-\bp$ increases rapidly. 
Our numerical results are well described by the function
\begin{equation}
  \label{babp_fit}
  (\ba-\bp)(g_0^2) = -0.00093\ g_0^2 \times {{1+23.3060\ g_0^2-27.3712\
  g_0^4}\over{1-0.9833\ g_0^2}},
\end{equation}
which incorporates the perturbative result for small values of
$g_0^2$.
In the range $0.8881 \leq g_0^2 \leq 1$
our data are represented with an absolute deviation smaller than $0.003$. 

\subsection{Result for $\bm$}

\begin{figure}[htbp]
  \begin{center}
    \includegraphics[width=12.5cm]{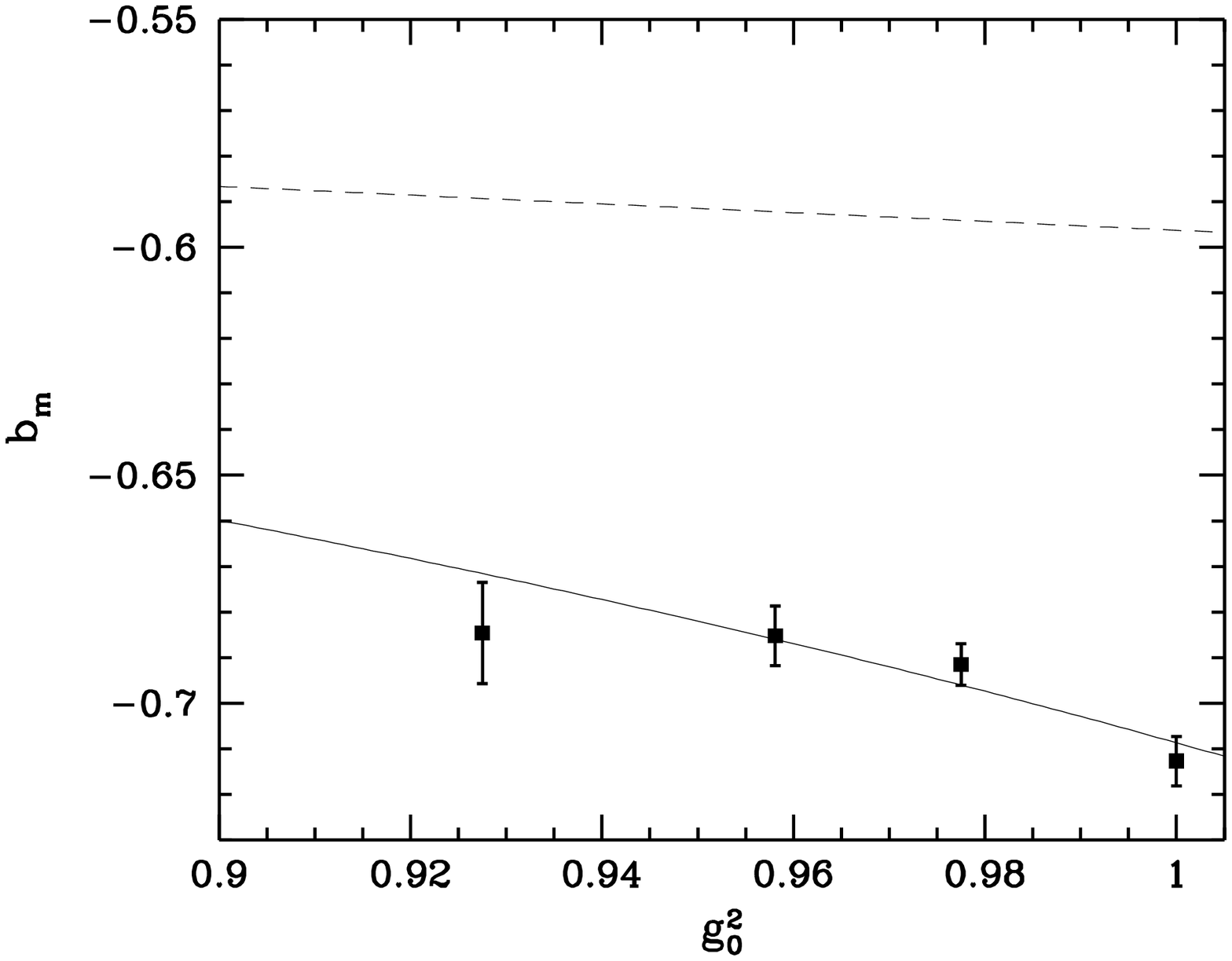}
    \caption{\footnotesize
      Results for $\bm$ as a function of $g_0^2$ from
      numerical simulations (filled squares) and from one-loop
      perturbation theory (dashed line).
      The full line represents the fit~(\ref{bm_fit}).
    \label{r_bm}}
  \end{center}
\end{figure}

The determination of the improvement
coefficient $\bm$ requires a subtle cancellation which is achieved by
setting the third bare mass parameter to the average 
of the other two~(\ref{eq:m3}). In terms of the hopping 
parameters this translates to
\begin{equation}
\label{eq:kappa3}
  \kappa_3 ={{2\kappa_1\kappa_2}\over{\kappa_1+\kappa_2}}.
\end{equation}
As we perform simulations with single precision arithmetic 
following the IEEE standard we expect relative roundoff errors 
of the size $6\times 10^{-8}$. This introduces a systematic error
which however can be estimated by repeating the derivation
of formula~(\ref{R_m}) and taking into account a small deviation of 
$\frac12(m_{0,1}+m_{0,2})-m_{0,3}$ from zero. 
This difference can be parametrized by the corresponding hopping
parameters which are the input parameters of the simulation
program. Thus we define 
\begin{equation}
  \label{eq:n4}
  \epsilon={{2\kappa_1\kappa_2}\over{\kappa_1+\kappa_2}} - \kappa_3.
\end{equation}
Then an additional term proportional to $\epsilon$ appears
in~(\ref{R_m}). The result is
\begin{equation}
  \label{eq:n3}
  R_{\rm m}^{\text{corrected}} = R_{\rm m} + \delta_{b_{\rm m}},\quad
  \delta_{b_{\rm m}}=2\epsilon{{(\kappa_1+\kappa_2)^2}\over{(\kappa_1-\kappa_2)^2}},
\end{equation}
up to terms of O($\epsilon^2$).
This effect is tabulated in table~\ref{tab:n3}. To obtain these values
we have used the single precision internal values for the hopping parameters.
\begin{table}[htbp]
  \begin{center}
    \begin{tabular}{|c||c|c|c|c|c|}\hline
      $L/a$          & 8& 10 & 12 & 16 & 24 \\ \hline
      $\delta_{b_{\rm m}}$ & $-0.0013$ & $-0.0021$ & $-0.0059$ & $-0.0135$ & $0.040$\\ \hline
    \end{tabular}
    \caption{\footnotesize Correction constant for $\bm$ as computed from
      equation~(\ref{eq:n3}).}
    \label{tab:n3}
  \end{center}
\end{table}
The correction constant turns out to be
negative for all lattice sizes except for $L/a=24$.
While on the smaller lattices it is rather small
it grows with the lattice size since the masses 
and thus $\kappa_1$ and $\kappa_2$ get closer to each other. 
We have decided to correct the statistically obtained
values of $\bm$ by adding $\delta_{\bm}$ for lattice sizes up to $L/a=16$.
At $L/a=24$ this effect is about five times larger than the target 
value for the statistical error of $b_{\rm m}$ so 
that we have chosen to omit this lattice in the discussion of $\bm$. 
As an estimate for the remaining systematic uncertainty 
we quote $\frac{1}{2}\delta_{\bm}$. Our final results for
$\bm$ are given in table~\ref{tab:n2} and shown in
figure~\ref{r_bm}. 
For large values of the coupling $g_0$ the value for $\bm$ decreases
rapidly. We find that the data are well described by the rational fit function
\begin{equation}
  \label{bm_fit}
  \bm(g_0^2) = 
       (-0.5-0.09623\ g_0^2)
   \times {{1-0.6905\ g_0^2+0.0584\ g_0^4}\over{1-0.6905\ g_0^2}},
\end{equation}
which asymptotically reproduces the perturbative
result~(\ref{bm_pert}). The fit describes the data with an absolute
deviation smaller than $0.013$.

\subsection{Alternative improvement conditions}

\begin{figure}[htb]
\begin{center}
\epsfig{figure=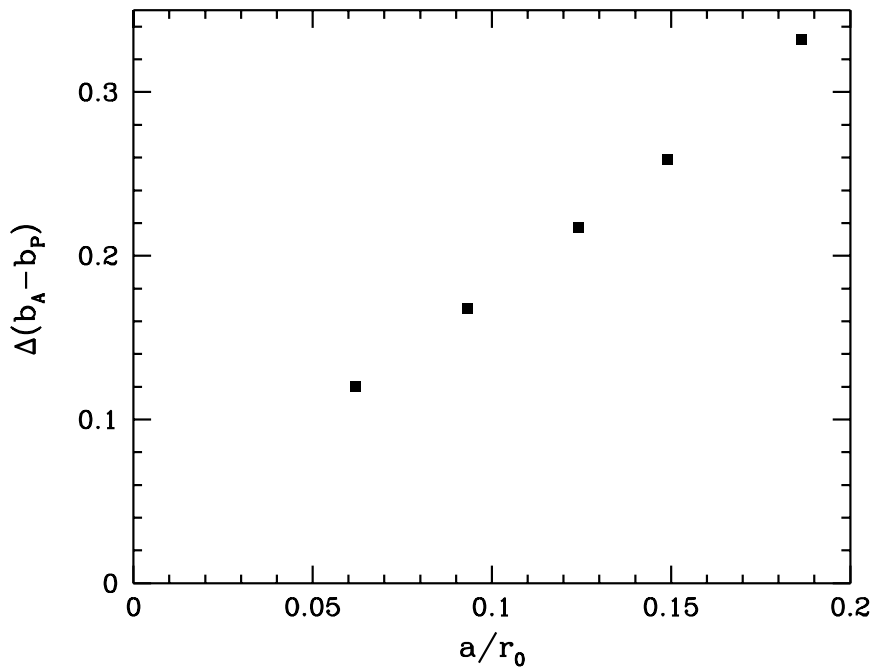, width=12.5 true cm}
\end{center} 
\vskip -1.0 true cm
\caption{\footnotesize
  The difference $\Delta(\ba-\bp)$ versus $a/r_0$. 
\label{delta_babp}}
\end{figure}

In order to verify the expectation that the difference
between two determinations of the improvement conditions
vanishes with a rate proportional to $a/r_0$ we now consider
a few examples. We start with $\ba-\bp$, where the choice
of the lattice derivative (improved vs. standard, cf. sect.~2)
made a big difference (cf.~fig.~\ref{babp_spread}). 
We thus define the difference
\begin{equation}
  \Delta(\ba-\bp) = \Rap(\text{impr. deriv.})-\Rap(\text{std. deriv.})
\end{equation}
and otherwise do not change the parameters. The
result is shown in fig.~\ref{delta_babp}. 
The behaviour is approximately linear in $a/r_0$.

The same difference for $\bm$ is rather small,
cp.~fig.~\ref{bm_spread}. Also alternative definitions of $\Delta \bm$
such as 
\begin{equation}
 \Delta \bm = R_{\rm m}(x_0=T/2)- R_{\rm m}(x_0=T/3).
\end{equation}
yield small values for the O($a$) ambiguities of $\bm$.
%
%

\subsection{Determination of $Z$}

For the determination of the renormalization constant in
the O($a$) improved theory we would like to keep the physics
 fixed up to errors of O($a^2$). Fortunately
this is already the case, for the smallness of $\ba-\bp$ 
implies that the renormalized O($a$) improved quark masses
are constant in units of $L$ with good numerical precision
(cp.~fig.~\ref{mass_tuning}).
\begin{figure}[htb]
\begin{center}
\epsfig{figure=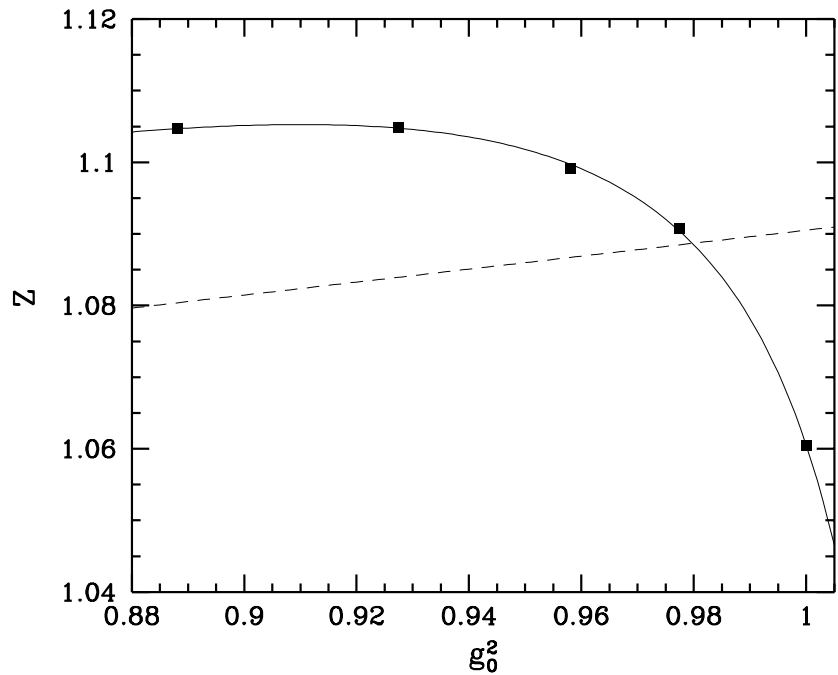, width=12.5 true cm}
\end{center} 
\vskip -1.0 true cm
\caption{\footnotesize
      Results for $Z$ as a function of $g_0^2$ from numerical
      simulations (filled squares) and 1-loop perturbation theory
      (dashed line). The solid line represents the fit
      function~(\ref{Z_fit}).
\label{r_Z}}
\end{figure}
\begin{figure}[htb]
\begin{center}
\epsfig{figure=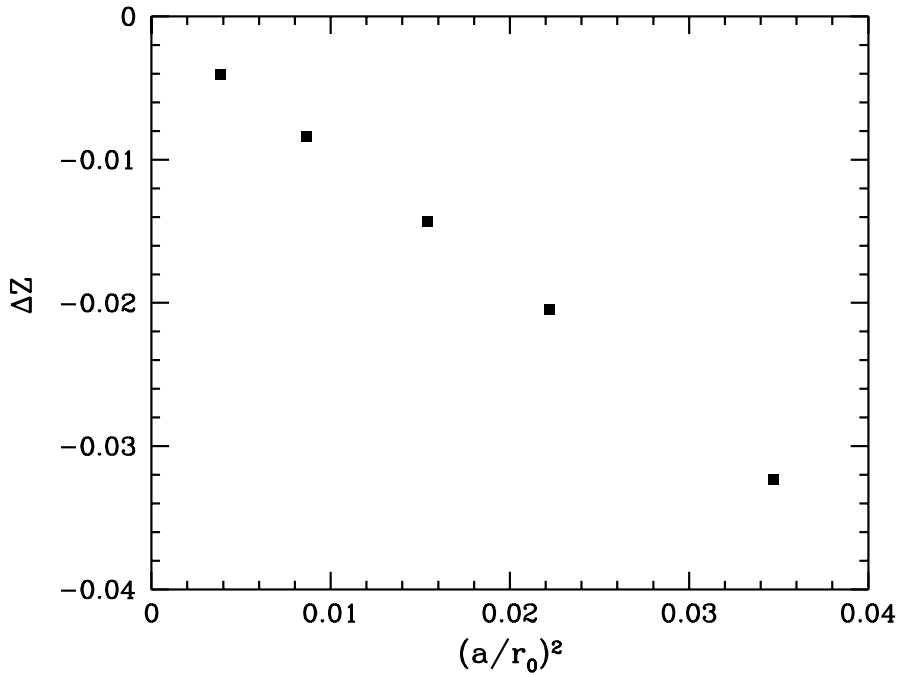, width=12.5 true cm}
\end{center} 
\vskip -1.0 true cm
\caption{\footnotesize
This plot shows the difference between the estimates
of $Z$ obtained from two alternative renormalization conditions.
\label{delta_Z}}
\end{figure}
To evaluate the estimator $R_Z$ one needs  
the combination of improvement coefficients 
$\ba-\bp-\bm$ [cp.~eq.~(\ref{R_Z})]. 
We here simply insert the value obtained with the 
fit functions~(\ref{babp_fit}) and~(\ref{bm_fit}). 
At the largest lattice size of our simulations $g_0$ is outside the range
of validity of~(\ref{bm_fit}). We use a linear fit to the last three data points 
for $b_{\rm m}$ to
extrapolate our data to this point. The resulting systematic error can
be estimated by calculating $Z$ with different values for $b_{\rm m}$. We
find that it is smaller than $10^{-4}$ and will neglect it in the
following. 
Our results for $Z$ are summarized in the fourth column of
table~\ref{tab:n2} and shown in figure~\ref{r_Z}.
In the range of couplings considered
we find that our data is represented by the fit function 
\begin{equation}
  \label{Z_fit}
  Z(g_0^2) = (1+0.090514\ g_0^2)\times{{1-0.9678\
  g_0^2+0.04284\ g_0^4 - 0.04373\ g_0^6}\over{1-0.9678\ g_0^2}},
\end{equation}
with a relative precision better than $0.04\%$. Again
the perturbative result~(\ref{Z_pert}) has been incorporated
into the fit.

It is important to check that another determination
of $Z$ differs by O($a^2$). In fig.~\ref{delta_Z} we plot
the difference between $R_Z$ determined with the improved
and with the standard lattice derivatives versus $a^2/r_0^2$
and see the expected roughly linear dependence.

\section{Summary and Conclusion}

We have carried out a detailed non-perturbative determination of 
the improvement coefficients $\bm$ and $\ba-\bp$ for
the range of $\beta$ values used in present large scale
simulations. In the case of $\ba-\bp$ we find that
the typical O($a$) ambiguity at $\beta=6.0$ is not small.
Rather than quoting a corresponding systematic error
we advocated the use of improvement conditions at
constant physics. This generalizes the
strategy previously applied to renormalization 
constants~\cite{paperIV}.
As a result, we have obtained the improvement coefficients
as smooth functions of the bare
coupling~[eqs.~(\ref{babp_fit}),(\ref{bm_fit})], 
and we checked that other choices of improvement
condition lead to differences which vanish with a rate proportional
to $a/r_0$. As a byproduct of our approach we also obtained
the finite renormalization constant $Z=\Zm\Zp/\Za$, eq.~(\ref{Z_fit}).
The same remarks as above apply to this case, except that the difference 
to other determinations vanishes with a rate proportional to $a^2/r_0^2$.
In contrast to the improvement coefficients, our result for 
$Z$ differs from 1-loop perturbation theory in the bare coupling
by no more than 3\%.

Imposing physical improvement conditions 
and the error analysis
are facilitated
by the use of the Schr\"odinger functional technology
and the use of simple direct estimators. 
It would be more difficult
to ensure the constant physics condition if smearing
techniques were necessary to enhance the signal, or if 
the improvement coefficients were obtained through a fitting procedure.
Nevertheless, where they overlap, 
our new results are compatible with those
of ref.~\cite{GiuliaRoberto} within the O($a$) and O($a^2$) ambiguities
discussed in this paper.
A drawback of the present method is that
in the case of $\bm$ we reached the limit of single precision
arithmetic, because the estimator of this coefficient relies
on the cancellation between bare quark masses which 
only happens up to rounding errors. As explained in detail,
this problem becomes more pronounced as the lattice 
size increases and prevented us from using our largest lattice
for an estimate of $\bm$.
We note that the strategy proposed here may be applied
in other cases and may certainly be combined with the
ideas of refs.~\cite{Gupta1,Gupta2} in order to determine
the coefficients $\ba$ and $\bp$ separately.

We end with some comments concerning the coefficient $\ca$
and the magnitude of the corrections associated with $\ba-\bp$ and
$\bm$.
Since the
original non-perturbative determination in ref.~\cite{paperIII}
the relatively large value of $\ca$ at $\beta=6.0$ (as compared
to one-loop perturbation theory) has been subject
to doubts by other authors~\cite{Gupta1,Gupta2}.
In these papers,
an alternative determination of $\ca$ was presented which yields
a numerical value which is roughly half as big at $\beta=6.0$.\footnote{
Some time ago, we had also considered alternative improvement conditions to 
determine $\ca$ \cite{MarcoRainer_notes}. In all cases when the lattice
artifact had a large sensitivity to $\ca$, the results were consistent with 
the original determination of ref.~\cite{paperIII}.}
One should note, however, that  all the scaling tests that have been carried
out for physical quantities~\cite{scalingI,scalingII} 
show the expected O($a$) improved continuum approach, albeit
with sizable O($a^2$) artifacts in some cases. 
Our present work contains additional examples 
of scaling. In fact, the quantities shown in figures 
\ref{delta_babp} and \ref{delta_Z}
have the advantage that they constitute pure lattice 
artifacts (rather than having an a priori unknown continuum limit). 
More precisely, $\Delta b_{\rm m}$
and $\Delta (b_{\rm A}-b_{\rm P})$ have to vanish in the limit
$a/r_0=0$ only when the theory is order $a$ improved, i.e. 
when $\csw$ and $\ca$ have the proper values. Similarly 
$\Delta Z=\rmO(a^2)$
is true only in this case. 
The convincing agreement with 
the expected behavior (cf.~figs. 
\ref{delta_babp} and \ref{delta_Z})
therefore constitutes new evidence that $\rmO(a)$ improvement
has been correctly implemented.

We have emphasized above, that in taking the continuum limit
of physical quantities the correct dependence of the improvement coefficients
on $g_0$ is crucial. In addition,
their overall magnitude determines whether they are 
relevant in practical applications. As we are considering coefficients
multiplying quark masses, this depends in particular on whether
one is interested in the physics of light quarks or heavy quarks.
In the range of bare couplings considered, the bare strange quark mass 
in lattice units is smaller than $0.04$ \cite{strange2}. 
Therefore, approximating
the $b$-coefficients determined here by 1-loop perturbation theory 
(as has been done in \cite{strange2})
does not introduce errors beyond the percent level for light quarks. 
On the other hand, for quark masses around the charm quark mass,
non-perturbative values should be used if one is interested in precisions
of a few percent. In addition, the difference $\Delta(\ba-\bp)$ translates 
into an effect of about 10\% for the situation of a D-meson \cite{UKQCD_FB}
at $\beta=6.0$  and about half as much at $\beta=6.2$. 
Unless one takes the continuum limit, this is
the magnitude of O$(a^2\mq)$ effects to be expected
for D-mesons in the improved theory and one may suspect 
O$(a^2\mq^2)$ terms to be even larger.

\vskip 1ex

This work is part of the ALPHA collaboration research programme.
We would like to thank G. de~Divitiis and M.~Kurth for collaboration
in the early stages of this work. Computer time on
the Quadrics machines at DESY-Zeuthen and 
the University of Rome ``Tor Vergata'' is gratefully acknowledged. 
S.~Sint acknowledges support by the European Commission under
grant No.~FMBICT972442. J.~Rolf acknowledges a postdoc fellowship
by Deutsche Forschungsgemeinschaft (Graduiertenkolleg GK271).

\vfill
\eject

\end{document}